\documentclass{aa}
\usepackage{graphicx,natbib}
\bibpunct{(}{)}{;}{a}{}{,}

\def\apj{ApJ}

\def\aap{A\&A}

\begin{document}
\sloppypar


\title{Solar source of energetic particles in interplanetary space during the 2006 December 13 event}

\author{C. Li \inst{1}, Y. Dai \inst{2,3}, J. -C. Vial \inst{2}, C.
J. Owen \inst{1}, S. A. Matthews \inst{1}, Y. H. Tang \inst{3}, C.
Fang \inst{3}, \and A. N. Fazakerley \inst{1}} \institute{Mullard
Space Science Laboratory, University College London, Dorking, Surrey
RH5 6NT, UK
\\ \email{cl2@mssl.ucl.ac.uk} \and Institut d'Astrophysique
Spatiale, Universit$\acute{\rm e}$ Paris-sud 11 and CNRS, Orsay
91405, France \and Department of Astronomy, Nanjing University,
Nanjing 210093, China}

\date{Received 5 March 2009 / Accepted 12 June 2009}

\authorrunning{Li et al.}
\titlerunning{Solar source of energetic particles in the 2006 December 13 event}

\abstract{An X3.4 solar flare and a fast halo coronal mass ejection
(CME) occurred on 2006 December 13, accompanied by a high flux of
energetic particles recorded both in near-Earth space and at ground
level. Our purpose is to provide evidence of flare acceleration in a
major solar energetic particle (SEP) event. We first present
observations from ACE/EPAM, GOES, and the Apatity neutron monitor.
It is found that the initial particle release time coincides with
the flare emission and that the spectrum becomes softer and the
anisotropy becomes weaker during particle injection, indicating that
the acceleration source changes from a confined coronal site to a
widespread interplanetary CME-driven shock. We then describe a
comprehensive study of the associated flare active region. By use of
imaging data from HINODE/SOT and SOHO/MDI magnetogram, we infer the
flare magnetic reconnection rate in the form of the magnetic flux
change rate. This correlates in time with the microwave emission,
indicating a physical link between the flare magnetic reconnection
and the acceleration of nonthermal particles. Combining radio
spectrograph data from Huairou/NOAC, Culgoora/IPS, Learmonth/RSTN,
and WAVES/WIND leads to a continuous and longlasting radio burst
extending from a few GHz down to several kHz. Based on the
photospheric vector magnetogram from Huairou/NOAC and the nonlinear
force free field (NFFF) reconstruction method, we derive the 3D
magnetic field configuration shortly after the eruption.
Furthermore, we also compute coronal field lines extending to a few
solar radii using a potential-field source-surface (PFSS) model.
Both the so-called type III-$l$ burst and the magnetic field
configuration suggest that open-field lines extend from the flare
active region into interplanetary space, allowing the accelerated
and charged particles escape. }

\maketitle


\section{Introduction}

Based on decades of observations, it has been widely accepted that
there are two classes of solar energetic particle (SEP) events (Cane
et al. 1986; Kallenrode et al. 1992). The first class, lasting for
hours, known as impulsive SEP events, are characterized by a
distinct enhancement of high-Z elemental abundance (Reames et al.
1994) that presumably arises from resonant wave-particle interaction
in the flare reconnection region (Roth $\&$ Temerin 1997; Miller
1997). The second class, lasting for days and known as gradual SEP
events, are more intense and geo-effective. Such gradual events,
so-called major events, are always associated with fast coronal mass
ejections (CMEs) that drive shocks capable of accelerating the
particles they encounter (Kahler et al. 1986; Cane et al. 1991).
However, fast CMEs are associated with flares, representing
different manifestations of the same magnetic energy release process
(Harrison 1995; Zhang et al. 2001; Wang et al. 2003). Both are
capable of accelerating particles that form part of an SEP event, so
which process dominates the particle injection remains enigmatic.

The present paradigm decouples the source of the particles from
flares in major events for two main reasons. First, the particle
release time generally occurs somewhat later than the flare soft
X-ray emission peak (Cliver et al. 1982; Kahler 1994). However, if
we consider the effect of interplanetary scattering, and compare it
to the flare nonthermal emissions (for instance hard X-ray,
gamma-ray, and microwave emissions), the particle release time could
still be consistent with the flare emission (Tang et al. 2006).
Another argument against flare acceleration in major events is the
belief that all flare-accelerated particles are trapped on closed
magnetic field lines beneath the CME and cannot escape to
interplanetary space (Reames 2002). However, Cane et al. (2002)
found that major events are always accompanied by continuous and
longlasting radio bursts from around 500 MHz down to less than 1 MHz
generated by flare-accelerated electrons as they propagate away from
the Sun. This indicates that there should be open-field lines
extending from beneath CMEs such that interplanetary particles
originating in flare regions might be expected in major events.

Observational results have been found to support the flare
acceleration of SEPs in major events. The time-intensity profiles of
the SEP events that occurred on 2000 July 14 and 2003 October 28
both displayed two injection peaks: an initial impulsive increase
corresponding to the flare eruption, followed by a gradual component
until the CME-driven shock arrived at Earth (Li et al. 2007a; Li et
al. 2007b). It thus appears that there are two populations of SEPs:
the prompt one displays an impulse-like increase with a high Fe/O
ratio, being flare-related; the delayed one has a slow intensity
rise, and a low Fe/O ratio, being CME-driven shock related (Cane et
al. 2003). Desai et al. (2004) examined five interplanetary shocks
associated with SEP events, and found that the Fe/O ratio increases
with increasing energy. This observation contradicts the
anticipation that the shock-accelerated Fe/O ratio will decrease
with increasing energy. Furthermore, Cane et al. (2007) found that
the Fe/O ratios in these events can decrease even when the shock is
traveling through an Fe-rich ambient plasma. This observation does
not support the proposal of shock acceleration of the suprathermal
ion population, which is used to explain the high-Z element
abundance in major events (Tylka et al. 2005).

Numerical models also suggest that mixed particle acceleration by
both flares and CME-driven shocks provide much better fits to the
in-situ observations. Li \& Zank (2005) found that the proton
intensity profile of the flare-shock-mixed case shows an initial
rapid increase, owing to the flare acceleration, and followed by a
plateau similar to that of a pure shock case. Furthermore,
Verkhoglyadova et al. (2008) showed a good agreement between the
mixed acceleration modeled spectra of heavy ions and the in-situ
particle data.

So the classical two-type classification of SEP events appears not
to be so clear-cut. In an event, both the flare and CME-driven shock
could accelerate particles. To determine their relative roles in
particle acceleration during different phases of an event is the
main goal of this work. Fortunately, the event that occurred on 2006
December 13, the last ground level enhancement (GLE) event of solar
cycle 23, gives us such an excellent opportunity.

This paper is organized as follows: Section 2 starts with an
overview of the instrumentations applied in this study. Section 3
presents the observational results in detail, including the dynamics
of SEPs (time history, spectrum and anisotropy) and a study of the
associated flare active region (magnetic reconnection rate and radio
dynamic spectra). In section 4, we use both the nonlinear force-free
field (NFFF) reconstruction method and the potential-field
source-surface (PFSS) model to obtain the coronal magnetic
configurations above the flare active region. In section 5, we
discuss interpretation of the above analysis. Section 6 is dedicated
to conclusions.

\section{Instrumentations}

In this study, neutron monitor (NM) and spacecraft particle data are
presented to help us interpret the SEP event, and multi-wavelength
observations are used to study the associated flare active region.

\subsection{Particle emission}

The proton data in the energy range of 15 -- 500 MeV, with a
temporal resolution of 1 minute, was obtained by Geostationary
Operational Environment Satellite (GOES) 11, which is in
geostationary orbit above the Pacific ocean. The one-minute-averaged
solar cosmic ray (SCR) intensity was obtained from the Apatity NM,
which is located at N67.57E33.40 and has a very low cut-off rigidity
of 0.63 GV. At the time of the event, the Apatity NM was ¡°viewing¡±
in the direction towards the Sun along the interplanetary magnetic
field (IMF) lines, so it was in a suitable position to detect the
direct prompt SCR intensity.

The electron data from the Electron, Proton, and Alpha Monitor
(EPAM) onboard Advanced Composition Explorer (ACE) was used in the
53 -- 103 keV and 42 -- 65 keV energy band with, respectively,
temporal resolutions of 12 s and 5-minute-averaged. ACE orbits the
L1 libration point which is a point of Earth-Sun gravitational
equilibrium about 1.5 million km upstream from Earth. The ACE/EPAM
is designed to make in-situ measurements of ions and electrons in
interplanetary space. It is composed of five detectors with either 4
or 8 sectors that divide the spin space into approximately equal
regions, give us the opportunity to investigate particles
anisotropy. Details are given in Gold et al. (1998). Using the
LEFS60 detector, which is one of the two Low-Energy Foil
Spectrometer systems (LEFS60 and LEFS150) that points in a sunward
direction, we obtain electron intensities in 8 angular sectors in
the 42 -- 65 keV energy band which allows us to determine the
anisotropy of interplanetary particles.

\subsection{Multi-wavelength emission}

The light curve of the flare active region and the post flare
loop-like structure were obtained from the Transition Region and
Coronal Explorer (TRACE, Handy et al. 1999) in the 195 {\AA}
passband. The hard X-ray (HXR) flux in 12 -- 25 keV energy band was
obtained from the Reuven Ramaty High Energy Solar Spectroscopic
Imager (RHESSI, Lin et al. 2002). Using the maximum entropy method
(MEM-Sato algorithm, Sato et al. 1999), we also reconstructed the
RHESSI HXR sources respectively in 12 -- 25 keV and 25 -- 50 keV
energy band.

The imaging data obtained from Hinode Solar Optical Telescope
(Hinode/SOT, Tsuneta et al. 2008) in the Ca II H line at 396.85 nm
were used to investigate the evolution of the chromospheric flare
ribbons, and the Michelson Doppler Imager (MDI, Scherrer et al.
1995) onboard the Solar and Heliospheric Observatory (SOHO) provided
the longitudinal magnetic field map. To investigate the coronal
large scale disturbance, the 195 {\AA} images of Extreme-Ultraviolet
Imaging Telescope (EIT, Delaboudini\`{e}re et al. 1995) on board the
SOHO were also used.

The vector magnetogram, the microwave radio dynamic spectrum ranging
from 2.6 -- 3.8 GHz, and the radio flux at 3.8 GHz were all obtained
from the Huairou Station of the National Astronomical Observatories
of China (Huairou/NAOC). The decimeter radio dynamic spectrum
ranging from 180 to 1800 MHz was obtained from Culgoora radio
observatory, which is one of the stations of IPS radio and space
services (IPS), it has a temporal resolution of 3 seconds. The
metric radio dynamic spectrum ranging from 25 to 180 MHz was
obtained from the solar radio spectrograph located at Learmonth in
Australia, which is one of the instruments of the Radio Solar
Telescope Network (RSTN). It has a temporal resolution of 3 seconds.
The decameter and hectometer (DH) radio dynamic spectra in the
frequency range of 20 kHz -- 13.875 MHz and with a temporal
resolution of 1 minute, was recorded by the WAVES experiment
(Bougeret et al. 1995) onboard the Wind spacecraft.

\section{Observations}

\subsection{Dynamics of SEPs}

On 2006 December 13, an X3.4 class flare occurred with a fast halo
CME, bathing the Earth in a high flux of SEPs and a strong
interplanetary shock. In this section, we present the dynamics of
SEPs during the major event.

\subsubsection{Time history}

\begin{figure}
\centering
\includegraphics[angle=0,width=6.5cm]{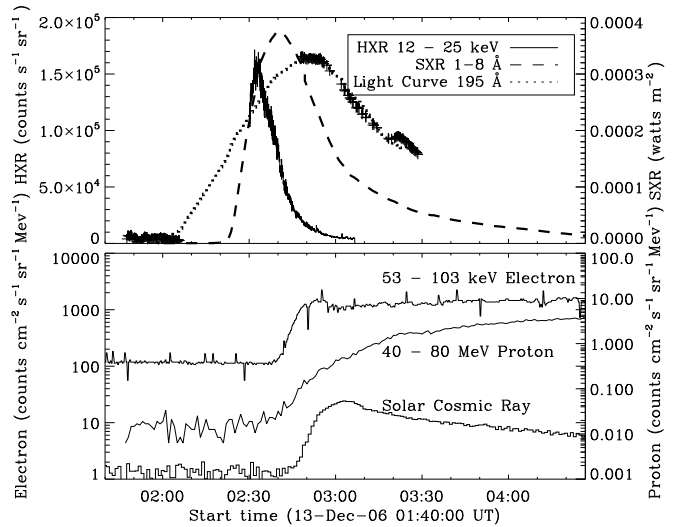}
\\~
\caption{Temporal profiles of electrons, protons, and SCRs, compared
with SXR emission, HXR emission, and light curve of the flare active
region. Upper panel: SXR flux ( 1 -- 8 $\rm {\AA}$, GOES), HXR flux
(12 -- 25 keV, RHESSI), and light curve of the active region (195
$\rm {\AA}$, TRACE) in arbitrary unit, plus symbols indicate the
data number (DN). Lower panel: electron intensity (53 -- 103 keV,
ACE/EPAM) with a temporal resolution of 12 s, one-minute-averaged
proton intensity (40 -- 80 MeV, GOES), and SCRs counts rate (Apatity
NM) in arbitrary unit.} \label{Fig1}
\end{figure}

Figure 1 shows the SXR, HXR, and TRACE light curve histories (upper
panel), together with the rise of electron, proton, and SCR count
rates (lower panel). This figure indicates that the SXR event in
GOES 1 -- 8 $\rm {\AA}$ started at 02:15 UT, peaked at 02:40 UT,
followed by a $\sim$ 2hr decay phase. The HXR emission in 12 -- 25
keV was observed by RHESSI from 02:29 UT, peaked at 02:32 UT, with a
decay phase lasting to 03:10 UT. The derivative of the SXR profile
peaked at 02:30 UT, which manifests as an approximate match in time
to the maximum of HXR emission. This consistent with expectations
from the Neupert effect (Neupert 1968) which describes the
relationship between the energy input by the nonthermal electrons
and the plasma's thermal response. The light curve of the flare
active region (X: [250, 500], Y: [-200, -50]) in unit of data number
(DN, plus symbols in the upper panel) obtained from TRACE 195 {\AA}
is also plotted. Although there was a data gap between 02:05 UT and
02:47 UT, we use a polynomial fit (short dashed line in the upper
panel) to the DN and this predicts the peak time as 02:50 UT.

The arrival of energetic particles are traced by ACE/EPAM, GOES-11,
and the Apatity NM (shown in the lower panel of Fig. 1). We assume
that energetic particles travel along the IMF lines at a speed of
$\upsilon$ with no scattering. Thus, with respect to the flare
emission time, we estimate the particle release time by subtracting
$\Delta t=\rm S/\upsilon-8.3\,\rm$ minutes from the observed time at
1 AU, where S is the length of IMF lines and $\upsilon$ is the
velocity of energetic particles. S $\sim$ 1.1 AU corresponding to a
solar wind velocity of about 700 km/s during this event. The
intensity onset is determined by 3$\sigma$ excess above background,
where $\sigma$ is the standard deviation of the particle count
rates. For GOES $\rm P_{5}$ (40 -- 80 MeV) and ACE electrons (53 --
103 keV), we take $\upsilon$ to be approximately 0.4$c$ and 0.6$c$,
where $c$ is the velocity of light. Then the evaluated proton and
electron release times are respectively 02:45 UT and 02:41 UT. For
Apatity NM SCRs, because of its low cut-off rigidity, we take the
associated proton energy as the atmospheric absorption cut-off
energy of 435 MeV and $\upsilon$ to be 0.7$c$. Then the evaluated
SCRs release time is 02:45 UT. Combining all the results and taking
the systematic errors into account, we obtain the SEPs release time
is 02:43 UT $\pm$ 3 minutes.

From Fig. 1, the comparison between the increase profiles of SEPs
and SXR emission, HXR emission, 195 $\rm {\AA}$ light curve of the
flare active region, and the above analysis, it is found that the
particle release time coincides with the flare emission, namely with
the rising phase of the 195 $\rm {\AA}$ light curve, the peak of the
SXR emission, and the decay phase of the HXR emission.

\subsubsection{Spectrum}

\begin{figure}
\centering
\includegraphics[angle=0,width=7.cm]{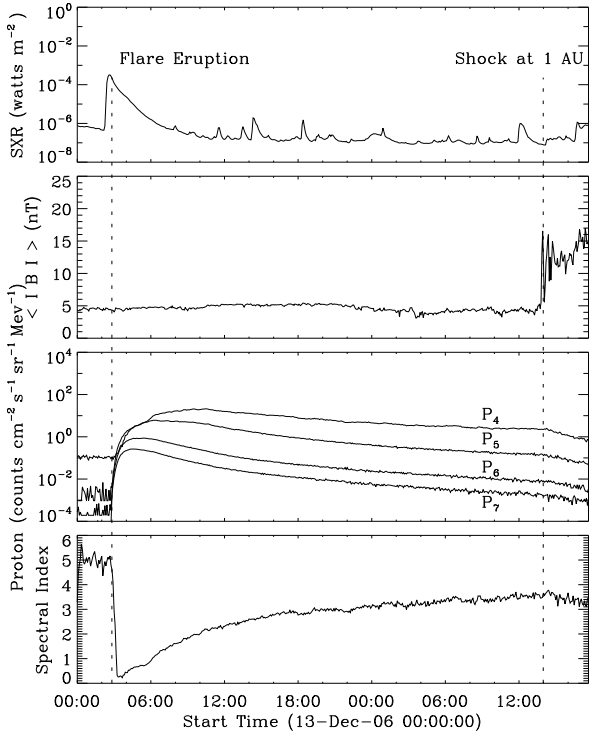}
\\~
\caption{From top to bottom: SXR flux (1 -- 8 $\rm {\AA}$, GOES),
magnetic field magnitude (ACE/MAG), proton intensity (15 -- 500 MeV,
GOES), and spectral index evaluated during the injection of protons.
Vertical dashed lines indicate the time of flare eruption and
CME-driven shock reaches 1 AU.} \label{Fig2}
\end{figure}

Figure 2, from top panel to bottom, respectively shows temporal
profiles of SXR flux, magnetic field magnitude, proton intensity,
and derived proton spectral index for a period of two days. It is
found that the injection profiles of protons have an impulsive
component during the flare emission, which is followed by a gradual
one. The latter lasts until the CME-driven shock reaches 1 AU at
14:00 UT on 14 December, as marked by vertical dashed lines, when
the magnetic field shows large increase, while the proton intensity
shows an obvious decrease.

Assuming a power law spectrum $f(E) \propto E^{-\gamma}$, and using
the different channel data from GOES in the energy range of 15 --
500 MeV, we get the proton spectral index during the particle
injection, which is shown in the bottom panel of Fig. 2. It shows a
soft background spectrum with an index $\gamma \sim 5$ before the
SEP event. This suddenly becomes obviously hard during the flare
eruption, then becomes increasely softer during the particle
injection until the CME-driven shock reaches 1 AU.

Furthermore, we present the spectra at three individual times in
Fig. 3. At the first peak flux, consistent with velocity dispersion
that results in lower energetic particles reaching the peak a bit
later than the higher ones, the spectral index $\gamma \sim 1.6$; at
18:00 UT 13 December, the spectral index $\gamma \sim 2.7$; at 14:00
UT 14 December, when the CME-driven shock reaches 1 AU, the spectral
index $\gamma \sim 3.6$. It is evident that the spectrum becomes
softer and softer during particle injection. Another interesting
phenomena to note is that the spectra at all three times appear to
become harder in high energy band, which are shown by the dashed
lines.

\begin{figure}
\centering
\includegraphics[angle=0,width=6.5cm]{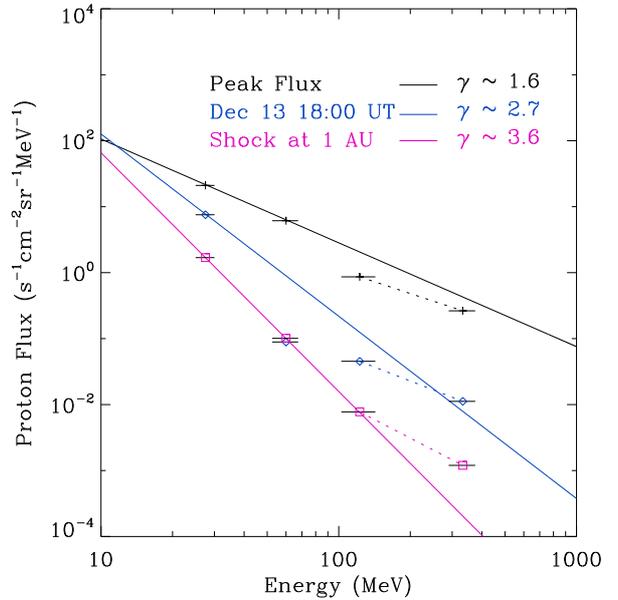}
\\~
\caption{The spectra at three different times during the SEP event.
The data are obtained from GOES with the energy range of 15 -- 500
MeV. Note that the spectrum becomes softer and softer during the
proton injection. Dashed lines indicate spectral hardening in high
energy band.} \label{Fig3}
\end{figure}

\subsubsection{Anisotropy}

In Fig. 4, three anisotropy pie plots derived from electron
intensities of 8 sectors of ACE/EPAM LEFS60 detector are shown
during the SEP event, each of them corresponds to the individual
time of the spectrum shown in Fig. 3. The large anisotropy at 03:00
UT 13 December is clearly seen in the left panel of Fig. 4, where
the electron count rates reach nearly 4000 c $\rm s^{-1}$ in sectors
3 and 4, smaller count rates are recorded by sectors 2 and 5. The
very weak anisotropy at 18:00 UT 13 December is shown in the middle
panel, where only sector 8 records obviously smaller count rates
than other sectors. From the above two panels, we can surmise the
accelerated particles direction of travel relative to the spacecraft
is from sectors 3 and 4 to sectors 7 and 8 during the first several
hours. The right panel shows the nearly isotropic distribution at
14:00 UT 14 December, indicating particles impact the spacecraft
from all the directions when the CME-driven shock reaches the
near-Earth space. Such a change in anisotropy suggests that particle
acceleration source shifts from a confined coronal site to a
widespread interplanetary region.

\subsection{Associated flare active region}

The source region of the major event was rooted in NOAA active
region 10930. When the flare occurred, the position of this region
was roughly at S06W24, which was near to the nominal well-connected
region or the foot-point of interplanetary magnetic field (IMF)
lines connecting the Sun to the Earth. In this section, we present
observations of the associated flare active region.

\begin{figure}
\centering
\includegraphics[angle=0,width=8.cm]{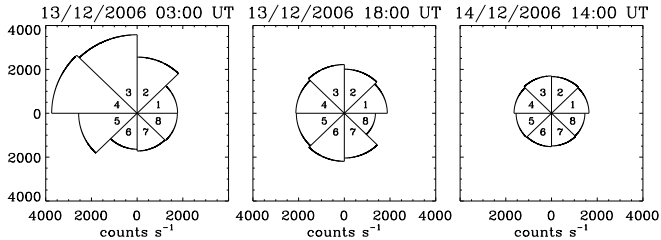}
\\~
\caption{Anisotropy plots from ACE/EPAM (42 -- 65 keV electron).
Left panel: strong anisotropy is seen in sectors 3 and 4 measured at
03:00 UT 13 December. Middle panel: weak anisotropy is only seen in
sector 8 at 18:00 UT 13 December. Right panel: nearly isotropic
distribution measured at 14:00 UT 14 December.} \label{Fig4}
\end{figure}

\subsubsection{Flare magnetic reconnection}

Figure 5 shows the post-flare loop-like structure in the high-pass
filtered TRACE $195\rm \AA$ image at 03:20:39 UT, which is overlaid
with MEM-Sato reconstructed RHESSI hard X-ray sources. Blue contour
lines indicate the 25 -- 50 keV nonthermal bremsstrahlung sources
integrated from 02:47:20 to 02:52:20 UT, and red contour lines
indicate 12 -- 25 keV thermal sources integrated from 02:42:20 to
02:47:20 UT. It is evident that the flare is a typical two-ribbon
system.

From the analytical (Forbes $\&$ Lin 2000) and numerical (Chen et
al. 1999) results, along with a large amount of two-ribbon flare
observations, it is accepted that the flare ribbon expansion in the
chromosphere is the signature of progressive magnetic reconnection
in the corona. The free energy contained in the non-potential
magnetic field can be rapidly dissipated into particle acceleration
and plasma heating at the reconnection region. How fast the magnetic
reconnection proceeds corresponds to how quickly the ribbons expand
and how strong the magnetic field is that the ribbons sweep through.
Thus we can estimate the magnetic reconnection rate in the form of
the magnetic flux change rate (Qiu et al. 2002):
\begin{equation}\label{1}
    \varphi_{rec}=\frac{\partial\Phi_{B}}{\partial t}=
    \frac{\partial}{\partial t}\int B_{n}dA,
\end{equation}
where $\Phi_{B}$ is the magnetic flux, $B_{n}$ the magnetic field
that the flare ribbons sweep through, and $dA$ the new area swept
through by the ribbons. It can be further compared with the
microwave emission to understand the role of the flare magnetic
reconnection in accelerating nonthermal particles.

Figure 6 shows the X3.4 flare observed by Hinode/SOT and the
SOHO/MDI magnetogram of the flare active region with the
trajectories of the two ribbons superposed. The Hinode/SOT Broadband
Filter Imager (BFI) produces images with broad spectral resolution
in 6 bands (CN band, Ca II H line, G-band, and 3 continuum bands) at
the highest spatial resolution, 0.0541 arcsec/pixel, over a 218"
$\times$ 109" field of view (FOV). The Ca II H line at 396.85 nm
image shows the chromospheric structure, so we use it to evaluate
the flare ribbon separation and get the new area $dA$ swept through
by the ribbons, the time cadence is $\sim$ 2 minutes during the
flare eruption. The magnetic field $B_{n}$ can be obtained from the
longitudinal component of the MDI magnetogram at 01:39:01 UT before
the flare eruption. Then the magnetic reconnection rate described in
formula 1 can be obtained.

The rate of magnetic flux change is evaluated for each of the two
ribbons, which respectively sweep through positive and negative
magnetic field, as $\varphi_{rec(+)}$ and $\varphi_{rec(-)}$,
$\varphi_{rec}$ is the average of them. In Fig. 7, the inferred
magnetic reconnection rate is shown in comparison with the microwave
emission at 3.8 GHz generated by high-speed electrons via
synchrotron. Both the radio flux of right-handed circular
polarization (RHCP) and left-handed circular polarization (LHCP) are
shown. Generally speaking, a good temporal correlation is found.
Furthermore, the correlation coefficient of $\varphi_{rec(+)}$,
$\varphi_{rec(-)}$, and $\varphi_{rec}$ with the RHCP radio flux is
calculated as 0.72, 0.80, and 0.87, with the LHCP radio flux as
0.84, 0.74, and 0.90. Such a good correlation indicates a strong
physical link between the flare magnetic reconnection and the energy
release, and the flare magnetic reconnection plays an important role
in accelerating nonthermal particles.

\begin{figure}
\centering
\includegraphics[angle=0,width=7.2cm]{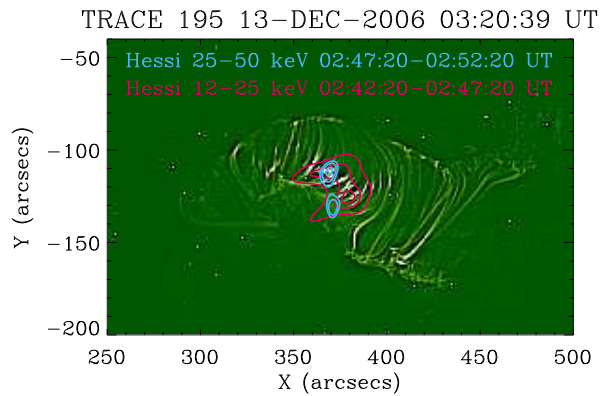}
\\~
\caption{Post flare loop-like structure shown by the high-pass
filtered TRACE $195\rm \AA$ image at 03:20:39 UT, overlaid by RHESSI
hard X-ray sources (Rotating modulation collimators 4 -- 8, with
MEM-Sato image reconstruction algorithm). Blue contour lines
indicate 25 -- 50 keV hard X-ray sources integrated from 02:47:20 to
02:52:20 UT, and red 12 -- 25 keV integrated from 02:42:20 to
02:47:20 UT.} \label{Fig5}
\end{figure}

\begin{figure}
\centering
\includegraphics[angle=0,width=8.cm]{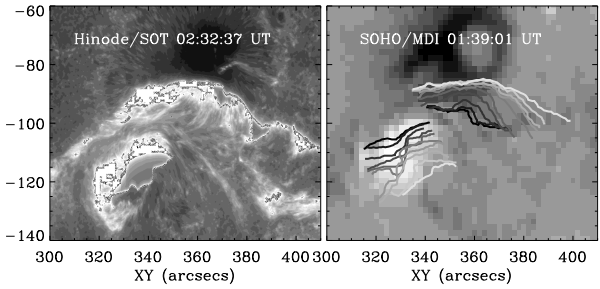}
\\~
\caption{Left panel: Snapshot of the X3.4 flare observed by
Hinode/SOT at 396.85 nm. Right panel: SOHO/MDI magnetogram of the
NOAA active region 10930 with the trajectories of the two ribbons
superposed. The color from dark to white indicates the separation
process of the ribbons.} \label{Fig6}
\end{figure}

\begin{figure}
\centering
\includegraphics[angle=0,width=7.3cm]{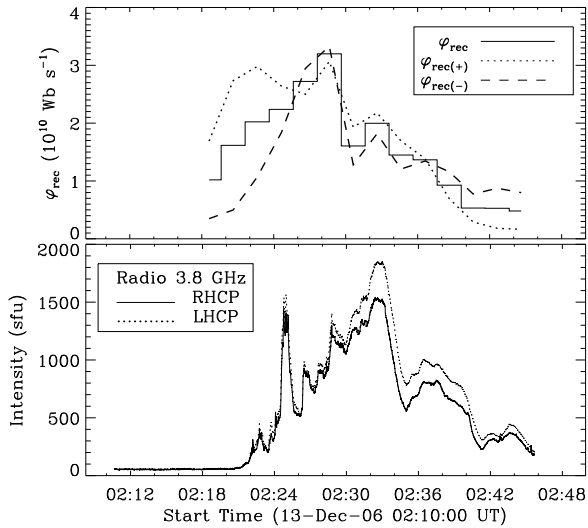}
\caption{Magnetic reconnection rate inferred from the X3.4
two-ribbon flare compared with the microwave radio emissions at 3.8
GHz. In the upper pannel, the magnetic reconnection rate is
evaluated for each of the two ribbons as $\varphi_{rec(+)}$ and
$\varphi_{rec(-)}$, $\varphi_{rec}$ is the mean value of the two. In
the lower panel, the solid line is the radio flux of right-handed
circular polarization (RHCP), dashed line of Left-handed (LHCP).}
\label{Fig7}
\end{figure}

\subsubsection{Radio dynamic spectra}

Solar radio bursts provide a clue to examine whether or not
particles accelerated in the flare site could escape and be observed
in near-Earth space. We combine various spectrographs to analyze the
radio dynamic spectra over an extended frequency range. Figure 8
shows a continuous and longlasting type III radio burst from $\sim$
1 GHz to 20 kHz, that starts at 02:24 UT and lasts about 20 minutes
in low frequency. This kind of type III burst, or so-called type
III-$l$ burst, shows characteristics of low-frequency extension and
longlasting duration (longer than the usual type III burst of 5 --
10 minutes). A type II radio burst starts at 02:27 UT with an
initial frequency of 180 MHz is also recorded.

Radio bursts (type I through V) are generated by electron streams as
they propagate along magnetic field lines from solar corona to
interplanetary medium (Wild et al. 1963). The emission frequency
corresponds to the local plasma frequency ($9\sqrt{n_{e}}$ kHz). In
detail, the plasma frequency of 1 GHz corresponds to $n_{e} \sim
10^{10} \ \rm cm ^{-3}$ and a coronal height of less than 0.1
$R_{s}$ above the photosphere, 20 kHz corresponds to $n_{e} \sim 10
\ \rm cm ^{-3}$ and the near-Earth plasma. Thus the type III-$l$
burst suggests the existence of open magnetic field lines from a low
coronal site, probably the flare active region, to the
interplanetary space.

An phenomena to note in Fig. 8 is that the type III-$l$ burst seems
to be an extension of the 2.6 -- 3.8 GHz microwave emission. This
suggests that the related electrons arise from the same flare
magnetic reconnection region, one population is trapped in closed
magnetic field and generates microwave emission through synchrotron,
another population escapes along opened magnetic field and generates
type III-$l$ burst through plasma emission.

\begin{figure}
\centering
\includegraphics[angle=0,width=6.8cm]{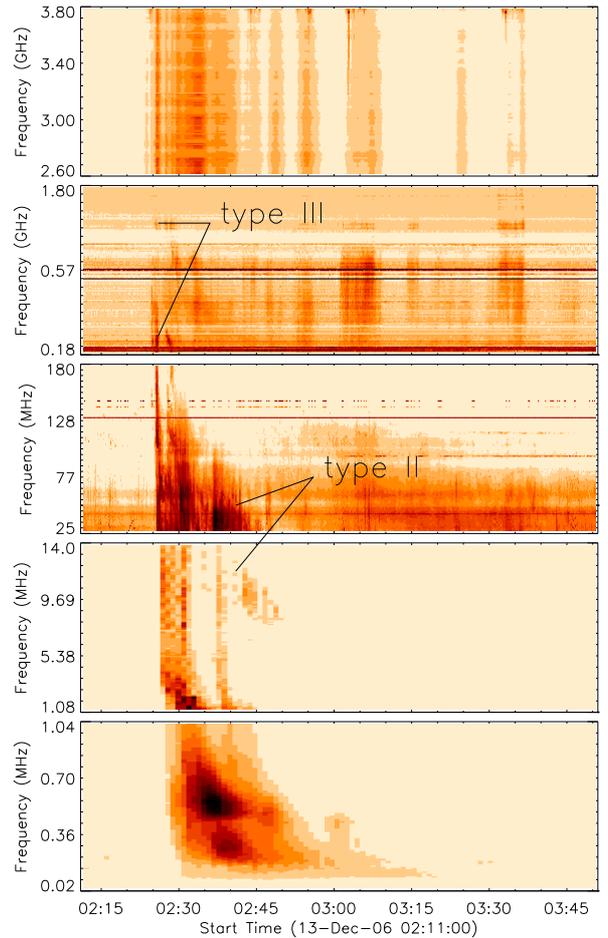}
\caption{From top to bottom: the radio dynamic spectra combining
Huairou/NAOC with the frequency range of 2.6 -- 3.8 GHz,
Culgoora/IPS of 180 to 1800 MHz, Learmonth/RSTN of 25 -- 180 MHz,
WAVES/WIND Rad2 1.075 MHz -- 13.875 MHz, and Rad1 of 20 kHz -- 1.04
MHz.} \label{Fig8}
\end{figure}

\section{Magnetic field modeling}

\begin{figure}
\centering
\includegraphics[angle=0,width=8.5cm]{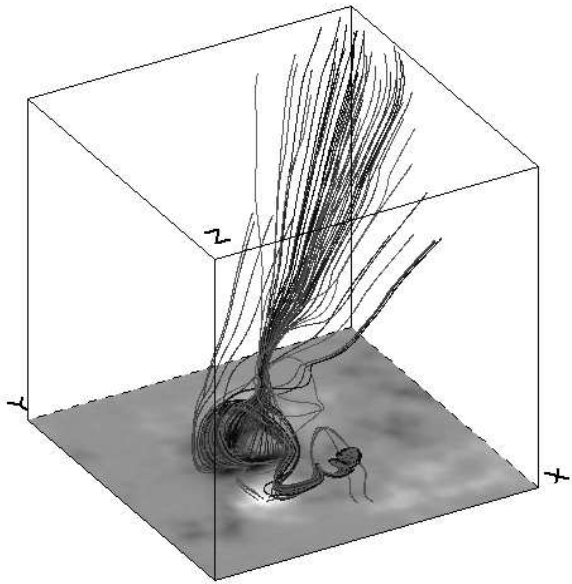}
\caption{Magnetic field configuration of the NFFF lines computed
based on the photospheric vector magnetogram of Huairou/NOAC. The
spatial domain of the magnetogram is $169^{"}\times169^{"}$, and the
calculated height is 122000 km. X and Y axes respectively indicate
west and north direction of the solar disk, and Z indicates radial
direction.} \label{Fig9}
\end{figure}

\begin{figure}
\centering
\includegraphics[angle=0,width=5.cm]{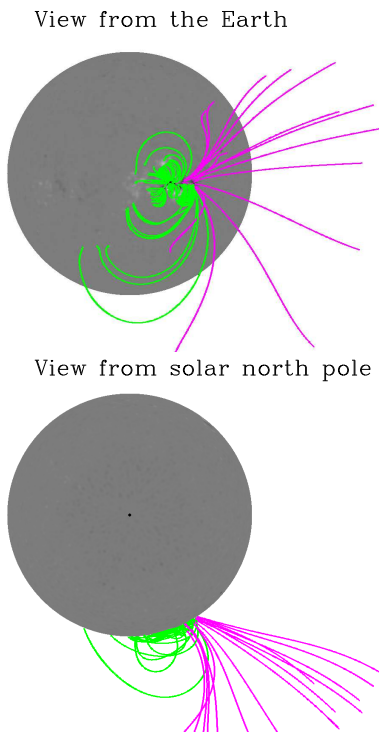}
\caption{Magnetic field configuration of the PFSS lines computed
based on the photospheric longitudinal magnetogram of SOHO/MDI. The
green lines mark the closed-field lines, and red lines mark the
open-field lines. Upper panel: the view direction is from the Earth.
Lower panel: the view direction is from the solar north pole.}
\label{Fig10}
\end{figure}

Another way to answer the question consists of examining the
magnetic field configuration above the flare active region. There is
currently no direct high-quality measurement of coronal magnetic
field. One method of obtaining the magnetic field configuration is
the precise extrapolation of the accurately observed photospheric
magnetic field. Song et al. (2006, 2007) developed a fast upward
integration NFFF reconstruction method. Instead of using finite
difference to express the basic NFFF partial derivatives, they
introduced smooth continuous functions to approach magnetic field
values, and a self-consistent compatibility condition for field
boundary values was considered.

Based on the photospheric vector magnetogram from Huairou/NAOC at
03:07 UT, just after the flare eruption, and the NFFF reconstruction
method, the 3D magnetic field configuration above the flare active
region is obtained and shown in Fig. 9. The magnetogram has magnetic
field noise levels of $\pm \ 10$ G for longitudinal fields and $\pm
\ 100$ G for transversal fields, the spatial resolution is
$2.8^{"}$, and the spatial domain is $169^{"}\times169^{"}$. The
magnetic field configuration is computed upwards to a height of
122000 km, 0.17 $R_{s}$ above the photosphere. It is evident that
the magnetic configuration shows non-potential complexity and
loop-like structure in the low corona, gradually becoming much
potential and then ``open" in the high corona above $\sim$ 0.1
$R_{s}$. Similar extrapolation results by various reconstruction
methods are also derived by Schrijver et al. (2008) and Guo et al.
(2008). Note that the ``open" lines mainly extend along the
northwest (XY direction) of the solar disk, in this direction
charged particles can easily access the nominal well-connected
region and then be released into interplanetary space.

Furthermore, we use the PFSS model developed by Schrijver \& DeRosa
(2003), which is available in the IDL-based solar software (SSW)
package, to identify the large-scale coronal magnetic configuration.
This model assumes a spherical source surface (where the magnetic
energy density equals to the plasma energy density). Inside of this
surface the magnetic field is derived from a potential Laplace
equation: $\nabla^{2}\phi$, outside which the magnetic field is
frozen into the solar wind (Schatten et al. 1969). It has been
successfully applied to confirm the existence of coronal magnetic
flux tubes, in which the type III radio sources originate (Klein et
al. 2008).

Based on the photospheric longitudinal magnetogram from SOHO/MDI at
00:04 UT and the PFSS code, the coronal magnetic field lines
extending to 2.5 $R_{s}$ are obtained and shown in Fig. 10. The
green lines mark the closed-field lines, and red lines mark the
open-field lines. The upper panel displays the view from the Earth,
with the top of the map indicating solar north and the right, solar
west. The lower panel displays the view from the solar north pole,
map bottom towards to the Earth. As in the NFFF model the open-field
lines, which are rooted in the active region, follow the westward
direction to the well-connection region and extend to interplanetary
space. Such open-field lines are a possible route for transporting
particles from the flare site.

\section{Discussion}

\subsection{Interpretation of particle release times}

In section 3.1.1, we find that there is no contradiction between the
particle injection and the flare emission. However, the evaluated
proton release time is $\sim$ 02:45 UT, which is a few minutes later
than the electron release at $\sim$ 02:41 UT. Similar delays are
also reported by Klein et al. (2005a) and Le et al. (2006) in the
SEP events occurred on 1999 June 29, 2000 May 01, and 2005 January
20.

One may suppose that the electron release coincides with the flare
eruption and is related to type III burst (Lin 1985), while protons
are accelerated in the subsequent CME-driven shock and related to
type II burst (Reames 1999). However, according to the observations
from Culgoora/IPS, the type II burst started at 02:27 UT, which is
much earlier than both of the proton and electron releases. Shock
acceleration has no species selectivity, so the question is why did
it not accelerate electrons and protons at the same time? Another
suggestion, introduced by Krucker $\&$ Lin (2000), is that protons
are released at a much higher altitude than electrons, and protons
with successively lower energies are released as the CME-driven
shock propagates to successively greater heights, corresponding to
the release times being successively delayed. However, from analysis
of the radio observations and the escaping deka-MeV protons, Klein
$\&$ Posner (2005b) found that the initial particles are accelerated
behind the front of the CME-driven shock. Moreover, it still cannot
explain the same release times between relativistic SCRs and lower
energetic protons in this event.

A reasonable possibility is that the interplanetary trajectories of
protons and electrons are different due to their distinct
pitch-angle scattering and mean free paths. For instance, Krucker
$\&$ Lin (2000) analyzed 26 events, and found that the derived
electron travel length is 1.1 -- 1.3 AU, but the proton travel
length is around 2 AU in 9 events. Using numerical models of
interplanetary transport to fit the in-situ observations from the
2000 July 14 event, Bieber et al. (2002) found that the electron and
proton mean free path are respectively 0.75 AU and 0.27 AU. Thus,
taking the interplanetary scattering and the mean free path of 0.08
-- 0.3 AU (Palmer 1982) into account, the proton release time could
be brought forward by a few minutes (Kahler 1994).

\subsection{Interpretation of particle spectra}

In section 3.1.2, we derive the proton spectra and find significant
change of the spectral index. For acceleration at a parallel shock,
particles gain energy by scattering freely between the converging
upstream and downstream plasma without influencing the shock
structure. The accelerated particle distribution function is a power
law form in momentum (Blandford and Ostriker 1978), $f(p)d^{3}p
p^{-\sigma}d^{3}p$, where $f(p)d^{3}p$ is the number of particles in
the volume $d^{3}p$ of momentum space per unit volume. The spectral
index $\gamma$ depends only on $r$, inverse of the compression
ratio, $\gamma=3r/(r-1)$, where $r=u_{1}/u_{2}$ and $u_{1}(u_{2})$
is the upstream (downstream) bulk plasma flow velocity. $r$ can be
determined from the Rankine-Hugonoit conservation relation. With
assumptions that the shock is planar (the gyro-radii of all
particles are small compared to the curvature of the shock), the gas
is ideal, and the adiabatic index 5/3, the relation can be reduced
to (Ferraro and Plumpton 1966; Ellison and Ramaty 1985):
\begin{equation}\label{2}
    Q_{1}r^{2}+(5Q_{1}+\frac{5}{3}M_{1}^{2})r-\frac{20}{3}M_{1}^{2}=0,
\end{equation}
where $Q_{1}$ is the upstream magnetic to gas pressure ratio and
$M_{1}$ is the upstream Mach number. For the coronal shock, taking
typical values of $T \sim 10^{6}$ K, $n \sim 10^{9}$ $\rm cm^{-3}$,
$B \sim $ 10 G, and the CME speed of 1774 km/s, we get $r \doteq
2.7$. For the interplanetary shock at 1 AU, taking observational
values from ACE data of $T \sim 10^{5}$ K, $n \sim 10$ $\rm
cm^{-3}$, $B \sim $ 5 nT, and $M_{1} \sim$ 8, we get $r \doteq 2.6$.

If particles are only accelerated by the CME-driven shock, we would
expect almost the same spectral index between the initial coronal
shock accelerated particles and the near-Earth interplanetary shock
accelerated particles. However, it is not the case, and the fact is
that the spectrum obviously becomes softer. In the current work, we
are not able to exclude the influence of energetic storm particles
(ESPs) during the passage of IP shocks, which are either accelerated
at the front shocks or trapped in the vicinity of shocks. Although
there are evidence that only ESPs with a few MeV or below show
obvious enhancement in this event (Liu et al. 2008), and that only
40$\%$ of 330 shock-related ESP events with energies greater than
1.5 MeV (Huttunen-Heikinmaa \& Valtonen 2009).

\begin{figure}
\centering
\includegraphics[angle=0,width=8.cm]{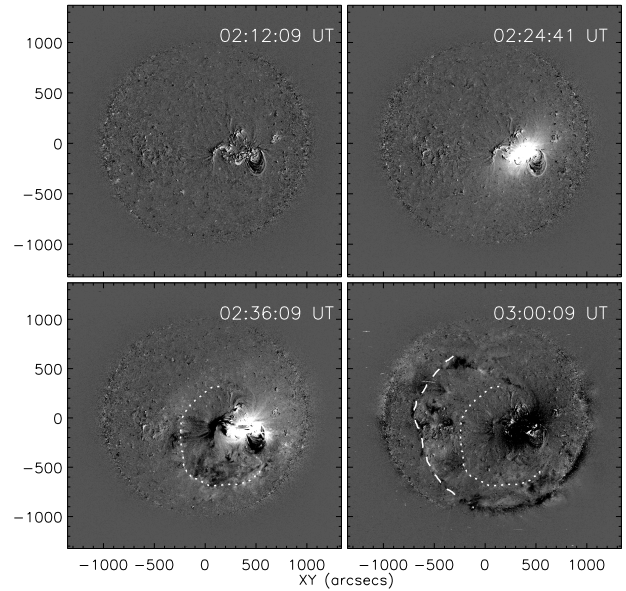}
\caption{The coronal disturbance on 2006 December 13. The four
images show SOHO/EIT 195 $\rm {\AA}$ ones at 02:12, 02:24, 02:36,
and 03:00 UT with a pre-event image subtracted from them,
respectively. Dashed lines indicate the fronts of the disturbance or
the EIT wave.} \label{Fig11}
\end{figure}

We cannot exclude an ideal model in which the acceleration shifts
from the coronal perpendicular shock to the interplanetary parallel
shock. The coronal transient wave or EIT wave might provide the
observational evidence. Krucker et al. (1999) found that the
electron release appeared related to the passage of EIT waves.
Torsti et al. (1999) found that the initial injection of deka-MeV
protons occurred during the period when the EIT wave was traversing
the western hemisphere of the Sun, which was interpreted as a moving
skirt of a quasi-perpendicular shock wave. However, recent studies
(Dalann\'{e}e 2000; Harra \& Sterling 2003; Chen et al. 2005;
Attrill et al. 2008; Dai et al. 2009) revealed that EIT waves are
not real material shocks, but coronal manifestations of the
expanding and opening magnetic fields. A large-scale coronal
disturbance or EIT wave was observed during this event. From the
running difference of EIT 195 $\rm {\AA}$ images (shown in Fig. 11),
it is found that a large amount of coronal material was ejected
around 02:36 UT or before, corresponding to the process from the
coronal brightening to dimming. This process may open or reconfigure
quite much of the magnetic field in the corona, allowing the
flare-accelerated particles to escape into interplanetary space.

We find that the proton spectra appear to become harder in high
energy band (shown by the dashed lines in Fig. 3). This is in
contradiction with the pure shock acceleration which would produce a
softer spectrum beyond 30 MeV (Ellison and Ramaty 1985), on the
other hand, is in favor to the contribution of flare acceleration
which is believed more prolific than the shock at producing high
energy particles (Tylka et al. 2005).

The unusual features of spectra, including spectral softening during
particle injection and spectral hardening in the high energy band,
give a reasonable possibility of the mixed particle acceleration by
both flare and the CME-driven shock.

\subsection{Interpretation of flare magnetic reconnection}

In section 4.1.1, we derive the magnetic reconnection rate in the
form of magnetic flux change rate. There still exists other forms,
such as continuous blue shift (Schmieder et al. 1987) and separation
velocity (Fletcher et al. 2004) of the flare ribbons. Both are
consistent with the concept in this study that the observed
signatures in chromosphere and the energy release in flares are due
to the magnetic reconnection at coronal site.

Assuming the separation of flare ribbons is translational symmetry,
and in a two-dimensional configuration, we can estimate the induced
electric field $E_{rec}$ along the current sheet in the magnetic
reconnection region by $E_{rec}=V_{\|}B_{n}$ (deduced from formula
1), where $V_{\|}$ is the horizontal velocity of the ribbon
expansion, $B_{n}$ the longitudinal magnetic field that the flare
ribbons sweep through. Given the magnetic field 650 G in the flare
active region, and the expansion velocity 14 km/s at the peak of
$\varphi_{rec}$, we get a maximum electric field of $\sim$ 10 V/cm.

Based on the DC electric field acceleration mechanism (Litvinenko
$\&$ Somov 1995), this value is effectively enough to accelerate
protons to a GLE typical energy of $\sim$ GeV in a time of $<$ 0.1
s. In this event, according to the registration of SCRs by middle
and low latitude neutron monitors, the maximum energy of the protons
is estimated to be $\sim$ 10 GeV (Karapetyan 2008). For the
electrons, based on the neutral beam current sheet theory (Martens
1988), the energy is always a factor $m_{e}/m_{p}$ less than proton
energy, and could reach the value of $\sim$ MeV. However, the DC
electric field acceleration mechanism could hardly produce the
high-Z elemental abundance, which probably arises from the
stochastic acceleration or the resonant wave-particle interaction.

\section{Conclusions}

This study contributes to the ongoing debate about which process,
flare or/and CME-driven shock, is responsible for particle injection
in major events (Tylka \& Lee 2006; Cane et al. 2007; Gopalswamy
2008). By use of particle data from near-Earth spacecraft and
ground-based neutron monitor, along with multi-wavelength
observations of the flare active region, we analyze the roles of
flare magnetic reconnection and CME-driven shock in accelerating
SEPs during the 2006 December 13 event. The key conclusions are as
follows:

\begin{enumerate}

\item The particle initial release time coincides with the flare
emission. The spectrum becomes softer and the anisotropy becomes
weaker during the particle injection, indicating the acceleration
source may change from a coronal site in the flare active region to
a widespread interplanetary CME-driven shock.

\item The inferred magnetic reconnection rate is temporally
well-correlated with the evolution of the microwave emission,
indicating a strong physical link between the flare magnetic
reconnection and the acceleration of nonthermal particles.

\item The continuous longlasting type III-$l$ radio burst and the
computed magnetic field configuration suggest the existence of open
magnetic field lines extending from the flare active region to
interplanetary space. Flare accelerated particles should thus be
present in the major event.

\end{enumerate}

\begin{acknowledgements} It is a pleasure to thank M. T. Song for
the use of his NFFF reconstruction method, and C. M. Tan, Y. Y. Deng
for their help with the Huairou/NAOC radio data. We are very
grateful to SOHO, TRACE, RHESSI, GOES, ACE, WIND, RSTN, IPS, and
Apatity NM for their open-policy data. Hinode is a Japanese mission
developed and launched by ISAS/JAXA, with NAOJ as domestic partner
and NASA and STFC (UK) as international partners. C. Li acknowledge
the financial assistance from Science $\&$ Technology Facilities
Council (STFC) of the UK.
\end{acknowledgements}


\begin{thebibliography}{}

\bibitem[\protect\citeauthoryear{}{}]{} Attrill, G., Harra, L. K.,
Van Driel-Gesztelyi, L., \& D\'{e}moulin, P. 2008, \apj, L101
\bibitem[\protect\citeauthoryear{}{}]{} Bieber, J. W., Dr\"{o}ge,
W., Evenson, P. A., et al. 2002, \apj, 567, 622
\bibitem[\protect\citeauthoryear{}{}]{} Blandford, R. D., \&
Ostriker, J. P. 1978, \apj, 267, L433
\bibitem[\protect\citeauthoryear{}{}]{} Bougeret, J. -L., Kaiser, M.
L., Kellogg, P. J., et al. 1995, Space Sci. Rev., 71, 5
\bibitem[\protect\citeauthoryear{}{}]{} Cane, H. V., McGuire, R. E., \&
von Rosenvinge, T. T. 1986, \apj, 301, 448
\bibitem[\protect\citeauthoryear{}{}]{} Cane, H. V., Reames, D. V., \&
von Rosenvinge, T. T. 1991, \apj, 373, 675
\bibitem[\protect\citeauthoryear{}{}]{} Cane, H. V., Erickson, W. C., \&
Prestage, N. P. 2002, J. Geophys. Res., 107, 1315
\bibitem[\protect\citeauthoryear{}{}]{} Cane, H. V., von Rosenvinge, T.
T., Cohen, C. M. S., \& Mewaldt, R. A. 2003, Geophys. Res. Lett.,
30, 8017
\bibitem[\protect\citeauthoryear{}{}]{} Cane, H. V., Richardson, I.
G., \& von Rosenvinge, T. T. 2007, Space Sci. Rev., 130, 301
\bibitem[\protect\citeauthoryear{}{}]{} Cliver, E. W., Kahler, S. W.,
Shea, M. A., \& Smart, D. F. 1982, \apj, 260, 362
\bibitem[\protect\citeauthoryear{}{}]{} Cliver, E. W., Kahler, S. W.,
\& Reames, E. V. 2004, \apj, 605, 902
\bibitem[\protect\citeauthoryear{}{}]{} Chen, P. F., Fang, C., Ding, M.
D., \& Tang, Y. H. 1999, \apj, 520, 853
\bibitem[\protect\citeauthoryear{}{}]{} Chen, P. F., Fang, C., \&
Shibata, K. 2005, \apj, 622, 1202
\bibitem[\protect\citeauthoryear{}{}]{} Dai, Y., Auch\`{e}re, F.,
Vial, J. -C., Tang, Y. H., \& Zong, W. G. 2009, ApJ, submitted
\bibitem[\protect\citeauthoryear{}{}]{} Delaboudini\`{e}re, J. -P.,
Artzner, G. E., Brunaud, J., et al. 1995, Sol. Phys., 162, 291
\bibitem[\protect\citeauthoryear{}{}]{} Delann\'{e}e, C. 2000, \apj,
545, 512
\bibitem[\protect\citeauthoryear{}{}]{} Desai, M. I., Mason, G. M.,
Wiedenbeck, M. E., et al. 2004, \apj, 611, 1156
\bibitem[\protect\citeauthoryear{}{}]{} Ellison, D. C., \& Ramaty,
R. 1985, \apj, 298, 400
\bibitem[\protect\citeauthoryear{}{}]{} Ferraro, V. C. A., \&
Plumpton, C. 1966, An Intruduction to Magneto-Fluid Mechanics (2d
ed; Oxford: Clarendon), p.101
\bibitem[\protect\citeauthoryear{}{}]{} Fletcher, L., Pollock, J.
A., \& Potts, H. E. 2004, Sol. Phys., 222, 279
\bibitem[\protect\citeauthoryear{}{}]{} Forbes, T. G., \& Lin, J.
2000, J. Atmos. Sol-Terr. Phys., 62, 1499
\bibitem[\protect\citeauthoryear{}{}]{} Gold, R. E., Krimigis, S. M.,
Hawkins, S. E., Haggerty, D. K., Lohr, D. A., Fiore, E., Armstrong,
T. P., Holland, G., \& Lanzerotti, L. J. 1998, Spa. Sci. Rev., 86,
541
\bibitem[\protect\citeauthoryear{}{}]{} Gopalswamy, N. 2008, 7th
Annual International Astrophysics Conference, AIP Conference
Proceedings, 1039, 196
\bibitem[\protect\citeauthoryear{}{}]{} Gosling, J. T. 1993, J.
Geophs. Res., 98, 18937
\bibitem[\protect\citeauthoryear{}{}]{} Guo, Y., Ding, M. D.,
Wiegelmann, T., \& Li, H. 2008, \apj, 679, 1629
\bibitem[\protect\citeauthoryear{}{}]{} Handy, B. N., Acton, L. W.,
Kankelborg, C. C., et al. 1999, Sol. Phys., 187, 229
\bibitem[\protect\citeauthoryear{}{}]{} Harra, L. K., \& Sterling,
A. C. 2003, ApJ, 587, 429
\bibitem[\protect\citeauthoryear{}{}]{} Harrison, R. A. 1995, \aap, 304,
585
\bibitem[\protect\citeauthoryear{}{}]{} Huttunen-Heikinmaa, K., \&
Valtonen, E. 2009, Ann. Geophys., 27, 767
\bibitem[\protect\citeauthoryear{}{}]{} Kahler, S. W., Cliver, E. W.,
Cane, H. V., McGuire, R. E., Stone, R. G. \& Sheeley, N. R. 1986,
Astrophys. J., 302, 504
\bibitem[\protect\citeauthoryear{}{}]{} Kahler, S. W. 1994, \apj, 428, 837
\bibitem[\protect\citeauthoryear{}{}]{} Kallenrode, S. W.,
Cliver, E. W., \& Wibberenz, G. 1992, \apj, 391, 370
\bibitem[\protect\citeauthoryear{}{}]{} Karapetyan, G. G. 2008,
Astroparticle Physics, 30, 234
\bibitem[\protect\citeauthoryear{}{}]{} Klein, K. -L., Krucker, S.,
Trottet, G., \& Hoang, S. 2005a, \aap, 431, 1047
\bibitem[\protect\citeauthoryear{}{}]{} Klein, K. -L., Posner, A.
2005b, \aap, 438, 1029
\bibitem[\protect\citeauthoryear{}{}]{} Klein, K. -L., Krucker, S.,
Lointier, G., \& Kerdraon, A. 2008, \aap, 486, 589
\bibitem[\protect\citeauthoryear{}{}]{} Krucker, S., Larson, D. E.,
Lin, R. P., \& Thompson, B. J. 1999, \apj, 519, 864
\bibitem[\protect\citeauthoryear{}{}]{} Krucker, S., \& Lin, R. P.
2000, \apj, 542, L61
\bibitem[\protect\citeauthoryear{}{}]{} Le, G. M., Tang, Y. H., \& Han, Y.
B. 2006, Chin. J. Astron. Astrophys., 6, 751
\bibitem[\protect\citeauthoryear{}{}]{} Li, C., Tang, Y. H., Dai, Y., Zong,
W. G., \& Fang, C. 2007a, \aap, 461, 1115
\bibitem[\protect\citeauthoryear{}{}]{} Li, C., Tang, Y. H., Dai, Y., Fang,
C., \& Vial, J. -C. 2007b, \aap, 472, 283
\bibitem[\protect\citeauthoryear{}{}]{} Li, G., \& Zank, G. P. 2005,
Geophys. Res. Lett., 32, 2101
\bibitem[\protect\citeauthoryear{}{}]{} Lin, R. P. 1985, Sol. Phys., 100, 537
\bibitem[\protect\citeauthoryear{}{}]{} Lin, R. P., Dennis, B. R.,
Hurford, G. J., et al. 2002, Sol. Phys., 210, 3
\bibitem[\protect\citeauthoryear{}{}]{} Litvinenko, Y. E., \&
Somov, B. V. 1995, Sol. Phys., 158, 317
\bibitem[\protect\citeauthoryear{}{}]{} Liu, Y., Luhmann, J. G.,
M\"{u}ller-Mellin, R., Schroeder, P. C., Wang, L., Lin, R. P., Bale,
S. D., Li, Y., Acuna, M. H., \& Sauvaud, J. -A. 2008, \apj, 689, 563
\bibitem[\protect\citeauthoryear{}{}]{} Miller, J. A. 1997, \apj, 491, 939
\bibitem[\protect\citeauthoryear{}{}]{} Neupert, W. M. 1968, \apj, 153, L59
\bibitem[\protect\citeauthoryear{}{}]{} Palmer, I. D. 1982, Rev. Geophys.
Space Phys., 20, 335
\bibitem[\protect\citeauthoryear{}{}]{} Qiu, J., Lee, J., Gary, D. E., \&
Wang, H. M. 2002, \apj, 565, 1335
\bibitem[\protect\citeauthoryear{}{}]{} Reames, D. V., Meyer, J. P., \&
von Rosenvinge, T. T. 1994, ApJS, 90, 649
\bibitem[\protect\citeauthoryear{}{}]{} Reames, D. V. 1999, Spa.
Sci. Rev., 90, 413
\bibitem[\protect\citeauthoryear{}{}]{} Reames, D. V. 2002, \apj, 571, 63
\bibitem[\protect\citeauthoryear{}{}]{} Roth, I., \& Temerin, M.
1997, \apj, 477, 940
\bibitem[\protect\citeauthoryear{}{}]{} Sato, J., Kosugi, T., \&
Makishima, K. 1999, PASJ, 51, 127
\bibitem[\protect\citeauthoryear{}{}]{} Schatten, K. H., Wilcox, J.
M., \& Ness, N. F. 1969, Sol. Phys., 6, 442
\bibitem[\protect\citeauthoryear{}{}]{} Scherrer, P. H., Bogart, R.
S., Bush, R. I., et al. 1995, Sol. Phys., 162, 129
\bibitem[\protect\citeauthoryear{}{}]{} Schmieder, B., Forbes, T.
G., Malherbe, J. M., \& Machado, M. E. 1987, \apj, 317, 956
\bibitem[\protect\citeauthoryear{}{}]{} Schrijver, C. J., \& DeRosa,
M. L. 2003, Sol. Phys., 212, 165
\bibitem[\protect\citeauthoryear{}{}]{} Schrijver, C. J., DeRosa,
M. L., Metcalf, T., et al. 2008, \apj, 675, 1637
\bibitem[\protect\citeauthoryear{}{}]{} Song, M. T., Fang, C., Tang, Y. H.,
Wu, S. T., \& Zhang, Y. A. 2006, \apj, 649, 1084
\bibitem[\protect\citeauthoryear{}{}]{} Song, M. T., Fang, C., Zhang, H.
Q., Tang, Y. H., Wu, S. T., \& Zhang, Y. A. 2007, \apj, 666, 491
\bibitem[\protect\citeauthoryear{}{}]{} Tang, Y. H., Li, C., \& Dai, Y.
2006, Proceedings of IAU Symposium, No. 233, 417
\bibitem[\protect\citeauthoryear{}{}]{} Torsti, J., Kocharov, L, G.,
Teittinen, M., \& Thompson, B. J. 1999, \apj, 510, 460
\bibitem[\protect\citeauthoryear{}{}]{} Tsuneta, S., Suematsu, Y.,
Ichimoto, K., et al. 2008, Sol. Phys., 249, 167
\bibitem[\protect\citeauthoryear{}{}]{} Tylka, A. J., Cohen, C. M.
S., Dietrich, W. F., Lee, M. A., Maclennan, C. G., Mewaldt, R. A.,
Ng, C. K., \& Reames, D. V. 2005, \apj, 625, 474
\bibitem[\protect\citeauthoryear{}{}]{} Tylka, A. J., \& Lee, M. A.
2006, \apj, 646, 1319
\bibitem[\protect\citeauthoryear{}{}]{} Verkhoglyadova, O. P.,
Li, G., Zank, G. P., \& Hu, Q. 2008, 7th Annual International
Astrophysics Conference, AIP Conference Proceedings, 1039, 214
\bibitem[\protect\citeauthoryear{}{}]{} Wang, H., Qiu, J., Jing, J., \&
Zhang, H. 2003, \apj, 593, 564
\bibitem[\protect\citeauthoryear{}{}]{} Wild, J. P., Smerd, S. F.,
\& Weiss, A. A. 1963, Anuu. Res. Aston. Astrophs., 1, 291
\bibitem[\protect\citeauthoryear{}{}]{} Zhang, J., Dere, K. P., Howard,
R. A., Kundu, M. R., \& White, S. M. 2001, \apj, 559, 452

\end{thebibliography}
\end{document}